\documentstyle[epsfig,12pt]{article}

\newcommand{\bce}{\begin{center}} 
\newcommand{\ece}{\end{center}}
\newcommand{\beq}{\begin{equation}}
\newcommand{\eeq}{\end{equation}}
\newcommand{\bea}{\vspace{0.25cm}\begin{eqnarray}}
\newcommand{\eea}{\end{eqnarray}}

\newcommand{\ba}{\begin{array}}
\newcommand{\ea}{\end{array}}

\newcommand{\doublespace}{
    \renewcommand{\baselinestretch}{1.6}\large\normalsize}

\def\lsim{\mathrel{\rlap{\lower4pt\hbox{\hskip1pt$\sim$}}
    \raise1pt\hbox{$<$}}}         
\def\gsim{\mathrel{\rlap{\lower4pt\hbox{\hskip1pt$\sim$}}
    \raise1pt\hbox{$>$}}}         

\def\Pom{{\bf I\!P}}

\def\beq{\begin{equation}}
\def\endeq{\end{equation}}
\def\arr{\begin{eqnarray}}
\def\endarr{\end{eqnarray}}
\makeindex

  \advance\oddsidemargin  by -1in
  \advance\evensidemargin by -1in
  \advance\topmargin      by -0.5in
\begin{document}

\vspace{2.0cm}

\begin{flushright}
ITEP(Ph)-2005-08
\end{flushright}

\vspace{1.0cm}

\begin{center}
{\Large \bf 
Charged currents, color dipoles and $xF_3$ at small $x$}.

\vspace{1.0cm}

{\large\bf R.~Fiore$^{1 \dagger}$ and V.R.~Zoller$^{2 \ddagger}$}

\vspace{1.0cm}

$^1${\it Dipartimento di Fisica,
Universit\`a     della Calabria\\
and\\
 Istituto Nazionale
di Fisica Nucleare, Gruppo collegato di Cosenza,\\
I-87036 Rende, Cosenza, Italy}\\
$^2${\it
ITEP, Moscow 117218, Russia\\}
\vspace{1.0cm}
{ \bf Abstract }\\
\end{center}
We develop the light-cone color dipole description of highly asymmetric
diffractive interactions of left-handed and right-handed electroweak bosons.  
We identify the origin  and estimate the strength of the left-right 
asymmetry effect in terms of the light-cone wave functions.
We  report an evaluation of the small-$x$ neutrino-nucleon DIS structure 
functions $xF_3$ and $2xF_1$ and present
comparison with experimental data.

\doublespace

\vskip 0.5cm \vfill $\begin{array}{ll}
^{\dagger}\mbox{{\it email address:}} & \mbox{fiore@cs.infn.it} \\
^{\ddagger}\mbox{{\it email address:}} & \mbox{zoller@itep.ru} \\
\end{array}$

\pagebreak



At small Bjorken $x$ the driving term of the inclusive/diffractive 
excitation  of charmed and
(anti)strange quarks in the charged current (CC) neutrino 
deep inelastic scattering (DIS)
is the $W^+$-gluon/Pomeron fusion, 
\beq
W^+g\to c\bar s \\
\label{eq:WG}
\eeq
and 
\beq
W^+\Pom\to c\bar s.
\label{eq:WP}
\eeq
Different aspects of
 the CC inclusive and diffractive 
DIS have been discussed in \cite{BGNPZ1,BGNPZ2}.

In the color dipole approach \cite{NZ91, M} 
(for the review see \cite{HEBECKER})  the
  small-$x$ DIS is treated in terms of the
interaction of the $c\bar s$ color dipole of size ${\bf r}$  
 with the target proton which is described by the beam- 
 and flavor-independent color dipole cross section
$\sigma(x,r)$. Once the light-cone wave function (LCWF) of a color dipole
state is specified the evaluation of  observable quantities
 becomes a routine quantum mechanical procedure.
In this communication we extend the color dipole analysis
 onto the CC  DIS  with particular emphasis on the left-right 
 asymmetry  of diffractive
interactions of electroweak bosons of different helicity.
We derive the relevant LCWF and evaluate the structure functions $xF_3$,
$\Delta xF_3$ and $2xF_1$. 
We  focus on the vacuum exchange dominated leading $\log(1/x)$ region
of $x\lsim 0.01$. 

At small $x$
the contribution of excitation of open 
charm/strangeness to the absorption cross section for scalar, $(\lambda=0)$,
left-handed, $(\lambda=-1)$,  and right-handed, $(\lambda=+1)$,  
$W$-boson of virtuality $Q^2$,
is given by the color dipole
factorization formula \cite{ZKL,BBGG}
\beq
\sigma_{\lambda}(x,Q^{2})
=\int dz d^{2}{\bf{r}} \sum_{\lambda_1,\lambda_2}
|\Psi_{\lambda}^{\lambda_1,\lambda_2}(z,{\bf{r}})|^{2} 
\sigma(x,r)\,.
\label{eq:FACTOR}
\eeq
In Eq.~(\ref{eq:FACTOR}) $\Psi_{\lambda}^{\lambda_1,\lambda_2}(z,{\bf{r}})$
 is the LCWF of
the $|c\bar s\rangle$ state with the $c$ quark 
carrying fraction $z$ of the $W^+$ light-cone momentum and 
$\bar s$ with momentum fraction $1-z$. The $c$- and $\bar s$-quark
helicities are  $\lambda_1=\pm 1/2$ and  $\lambda_2=\pm 1/2$, respectively.
The $W^+\to c\bar s$-transition vertex is specified as follows:
$$gU_{cs}\bar c\gamma_{\mu}(1-\gamma_5)s, $$
where $U_{cs}$ is an element of the CKM-matrix and the weak charge $g$ is
related to the Fermi coupling constant $G_F$,
\beq
{G_F\over \sqrt{2}}={g^2\over m^2_{W}}.
\label{eq:GF}
\eeq
 The polarization states of W-boson 
 carrying the laboratory frame four-momentum 
\beq
q=(\nu,0,0,\sqrt{\nu^2+Q^2}) 
\label{eq:Q}
\eeq
are described 
by the four-vectors $e_{\lambda}$, with
\bea
e_0={1\over Q}(\sqrt{\nu^2+Q^2},0,0,\nu)~, \nonumber\\
e_{\pm}=\mp{1\over \sqrt{2}}(0,1,\pm i,0)~, 
\label{eq:WPOL}
\eea
the unit vectors $\vec e_x$ and $\vec e_y$ being in $q_x$- and $q_y$-direction,
respectively. We find it convenient to use the basis of helicity spinors of 
Ref.~\cite{LB}.
 Then,  vector $(V)$ and axial-vector $(A)$ components of the LCWF
\beq
\Psi_{\lambda}^{\lambda_1,\lambda_2}(z,{\bf r})=
V_{\lambda}^{\lambda_1,\lambda_2}(z,{\bf r})  -
A_{\lambda}^{\lambda_1,\lambda_2}(z,{\bf r})
\label{eq:PSIVA}
\eeq
are as follows:
\bea
V_0^{\lambda_1,\lambda_2}(z,{\bf r})={\sqrt{\alpha_W N_c}\over 2\pi Q}
\left\{
\delta_{\lambda_1,-\lambda_2}\left[
2Q^2z(1-z)
\right.
\right.
 \nonumber\\
\left.
\left.
+(m-\mu)[(1-z)m-z\mu]\right]K_0(\varepsilon r)
\right.
 \nonumber\\
\left.
-i\delta_{\lambda_1,\lambda_2}(2\lambda_1)e^{-i2\lambda_1\phi}(m-\mu)
\varepsilon K_1(\varepsilon r)\right\}\,,
\label{eq:V0}
\eea
\bea
A_0^{\lambda_1,\lambda_2}(z,{\bf r})={\sqrt{\alpha_W N_c}\over 2\pi Q}
\left\{
\delta_{\lambda_1,-\lambda_2}(2\lambda_1)\left[
2Q^2z(1-z)
\right.
\right.
 \nonumber\\
\left.
\left.
+(m+\mu)[(1-z)m+z\mu]\right]K_0(\varepsilon r)
\right.
 \nonumber\\
\left.
+i\delta_{\lambda_1,\lambda_2}e^{-i2\lambda_1\phi}(m+\mu)
\varepsilon K_1(\varepsilon r)\right\}\,.
\label{eq:A0}
\eea
If $\lambda=\pm 1$
\bea
V_{\lambda}^{\lambda_1,\lambda_2}(z,{\bf r})=
-{\sqrt{2\alpha_W N_c}\over 2\pi }\left\{
\delta_{\lambda_1,\lambda_2}\delta_{\lambda,2\lambda_1}[(1-z)m+z\mu]
 K_0(\varepsilon r)
\right.
\nonumber\\
\left.
-i(2\lambda_1)\delta_{\lambda_1,-\lambda_2}e^{i\lambda\phi}
\left[(1-z)\delta_{\lambda,-2\lambda_1}+z\delta_{\lambda,2\lambda_1} \right]
\varepsilon K_1(\varepsilon r)\right\}\,,
\label{eq:VRL}
\eea
\bea
A_{\lambda}^{\lambda_1,\lambda_2}(z,{\bf r})=
{\sqrt{2\alpha_W N_c}\over 2\pi }\left\{
\delta_{\lambda_1,\lambda_2}\delta_{\lambda,2\lambda_1}(2\lambda_1)
[(1-z)m-z\mu]
 K_0(\varepsilon r)
\right.
\nonumber\\
\left.
+i\delta_{\lambda_1,-\lambda_2}e^{i\lambda\phi}
\left[(1-z)\delta_{\lambda,-2\lambda_1}+z\delta_{\lambda,2\lambda_1} \right]
\varepsilon K_1(\varepsilon r)\right\}\,,
\label{eq:ARL}
\eea
where
\beq
\varepsilon^2=z(1-z)Q^2+(1-z)m^2+z\mu^2
\label{eq:VAREPS}
\eeq
 and   $K_{\nu}(x)$ is the modified Bessel function. 
We do not consider Cabibbo-suppressed transitions and
$$\alpha_W={g^2/4\pi}.$$ 
The quark 
and antiquark masses are $m$ and $\mu$, respectively. The 
 azimuthal angle of ${\bf r}$ is denoted by  $\phi$.
To switch $W^+\to W^{-}$ one should perform the replacement
$m\leftrightarrow\mu $
in the equations above.

 The diagonal elements of  density matrix 
\beq
\rho_{\lambda\lambda^{\prime}}
=\sum_{\lambda_1,\lambda_2}\Psi_{\lambda}^{\lambda_1,\lambda_2}
\left(\Psi_{\lambda^{\prime}}^{\lambda_1,\lambda_2}\right)^*
\label{eq:RHO}
\eeq
 entering Eq.~
(\ref{eq:FACTOR}) are as follows:
\bea
\rho_{00}(z,{\bf r})
=\sum_{\lambda_1,\lambda_2}\left(\left|V_0^{\lambda_1,\lambda_2}\right|^2
+\left|A_0^{\lambda_1,\lambda_2}\right|^2\right)\nonumber\\
={{2\alpha_W N_c}\over (2\pi)^2 Q^2}\left\{
\left[\left[2Q^2z(1-z)
+(m-\mu)[(1-z)m-z\mu]\right]^2
\right.
\right.
\nonumber\\
\left.
\left.
+\left[2Q^2z(1-z)
+(m+\mu)[(1-z)m+z\mu]\right]^2\right]
\right.
\nonumber\\
\left.
\times K_0(\varepsilon r)^2 +[(m-\mu)^2 +(m+\mu)^2]
\varepsilon^2 K_1(\varepsilon r)^2\right\}
\label{eq:RHOSS}
\eea
and for $\lambda=\lambda^{\prime}=\pm 1$
\bea
\rho_{+1+1}(z,{\bf r})=\left|\Psi_{+1}^{+1/2+1/2}\right|^2
+\left|\Psi_{+1}^{-1/2+1/2}\right|^2\nonumber\\
={{8\alpha_W N_c}\over (2\pi)^2}(1-z)^2\left[m^2 K_0(\varepsilon r)^2
+\varepsilon^2 K_1(\varepsilon r)^2\right],
\label{eq:RHOR}
\eea
\bea
\rho_{-1-1}(z,{\bf r})=\left|\Psi_{-1}^{-1/2-1/2}\right|^2
+\left|\Psi_{-1}^{-1/2+1/2}\right|^2\nonumber\\
={{8\alpha_W N_c}\over (2\pi)^2}z^2\left[\mu^2 K_0(\varepsilon r)^2
+\varepsilon^2 K_1(\varepsilon r)^2\right].
\label{eq:RHOL}
\eea
 At $Q^2\to 0$ the terms $\sim m^2/Q^2, \mu^2/Q^2$ in Eq.~(\ref{eq:RHOSS})
remind us that $W$ interacts with the current which is not conserved
while the S-wave terms in Eqs.~(\ref{eq:RHOR}) and (\ref{eq:RHOL}) 
proportional to 
$m^2$ and  $\mu^2$  
 remind us that this current is 
the parity violating
 $(V-A)$-current.

The density of 
quark-antiquark $c\bar s$ states in the transversely polarized $W$-boson is 
\bea
\rho_{TT}={1\over 2}\left(\rho_{+1+1}+ \rho_{-1-1}\right)\nonumber\\
={{4\alpha_W N_c}\over (2\pi)^2}\left\{
\left[(1-z)^2m^2+ z^2 \mu^2\right]K_0(\varepsilon r)^2
\right.
\nonumber\\
\left.
+\left[(1-z)^2+ z^2\right]
\varepsilon^2 K_1(\varepsilon r)^2\right\}~.
\label{eq:RHOTT2}
\eea
One can see that our $\rho_{00}$ and $\rho_{TT}$  coincide with 
the probability densities  $|\Psi_L|^2$ and 
 $|\Psi_T|^2$  of Ref.~\cite{BGNPZ1} (see also Ref.~\cite{Z} where $z$-dependence 
of transverse and longitudinal CC cross sections has been discussed) .

        The momentum partition asymmetry of both  
$\rho_{-1-1}$ and $\rho_{+1+1}$ is striking,
the left-handed quark in  the decay of  left-handed $W^+$ gets the 
lion's share of the $W^+$ light-cone momentum.
The  nature of this  phenomenon  is very close to the nature of 
well known spin-spin correlations in the neutron $\beta$-decay.  
The observable which is 
  strongly affected by this left-right asymmetry  is the 
structure function of the neutrino-nucleon DIS named $F_3$. Its definition
in terms of $\sigma_{R}$ and $\sigma_{L}$ of Eq.~(\ref{eq:FACTOR}) is as follows:
\beq
2xF_3(x,Q^2)={Q^2\over 4\pi^2\alpha_{W}}
\left[\sigma_{L}(x,Q^{2})-\sigma_{R}(x,Q^{2})\right].
\label{eq:F3}
\eeq

To estimate consequences of the left-right asymmetry for
$F_3$ at high $Q^2$, such that 
\beq
{m^2\over Q^2}\ll 1,\,\,{\mu^2\over Q^2}\ll 1~,  
\label{eq:DLLA}
\eeq
one should take into account that
 the dipole cross-section $\sigma(x,r)$ in 
Eq.~(\ref{eq:FACTOR}) is related to the un-integrated
gluon structure function
${{\cal F}(x,\kappa^2)}={\partial G(x,\kappa^2)/\partial\log{\kappa^2}},$
 as follows \cite{FACTOR}:
\bea
\sigma(x,r)={\pi^2 \over N_c}r^2\alpha_S(r^2)
\int{d\kappa^2\kappa^2\over (\kappa^2+\mu_G^2)^2}
{4[1-J_0(\kappa r)]\over \kappa^2r^2}{{\cal F}(x_g,\kappa^2)}~.
\label{eq:SIGMA}
\eea
In the Double Leading Logarithm Approximation (DLLA), 
i.e. for small dipoles,
\bea
\sigma(x,r)\approx {\pi^2\over N_c} r^2\alpha_S(r^2)
G(x_g,A/r^2),
\label{eq:SMALL}
\eea
where $\mu_G=1/R_c$ is the inverse correlation radius of perturbative gluons
and $A\simeq 10$ comes from properties of the Bessel function $J_0(y)$.
Because of scaling violation $G(x,Q^2)$  rises with $Q^2$, but the product
$\alpha_S(r^2)G(x,A/r^2)$ is approximately flat in $r^2$.
At large $Q^2$ the leading
contribution to $\sigma_{\lambda}(x,Q^{2})$ comes from
the P-wave term, $\varepsilon^2 K_1(\varepsilon r)^2$, in
Eqs.~(\ref{eq:RHOR}) and (\ref{eq:RHOL}).
 The asymptotic behavior of the 
Bessel function, $K_1(x)\simeq \exp(-x)/\sqrt{2\pi/x}$ makes the 
$\bf{r} $-integration rapidly convergent at $\varepsilon r > 1$.
 Integration over $\bf{r}$ in Eq.~(\ref{eq:FACTOR}) yields
\beq
\sigma_L\propto \int_0^1 dz {z^2\over \varepsilon^2}\alpha_SG\sim 
{\alpha_SG\over Q^2}\log{Q^2\over \mu^2}
\label{eq:SIGL}
\eeq
and similarly 
\beq
\sigma_R\propto \int_0^1 dz {(1-z)^2\over \varepsilon^2}\alpha_SG
\sim  {\alpha_SG\over Q^2}\log{Q^2\over m^2}.
\label{eq:SIGR}
\eeq
The left-right asymmetry   certainly  affects also
 the slowly varying  product $\alpha_S G$ which for the purpose of
 crude estimate is taken at some rescaled
virtuality $\sim Q^2$ which is approximately/logarithmically  the same for 
$\sigma_L$ and $\sigma_R$. Hence,
\beq
\sigma_L-\sigma_R\propto 
{\alpha_SG\over Q^2}\log{m^2\over \mu^2}.
\label{eq:SIGLR}
\eeq 
Notice that in spite of the apparent asymmetry of the $z$-distribution
both $\sigma_L$ and $\sigma_R$ get equal scaling contributions from 
the integration domains near by the peaks  $z=1$ and $z=0$, respectively. 
Therefore, $xF_3$ is free of the end-point contributions.

At $Q^2\to 0$ and $\mu^2/m^2\ll 1$ the cross sections 
$\sigma_L$ and $\sigma_R$ are as follows:
\beq
\sigma_L\propto  
{\alpha_SG\over m^2}\log{m^2\over \mu^2}, \,\, 
\sigma_R\propto  
{\alpha_SG\over m^2}.
\label{eq:SIGLS}
\eeq

\begin{figure}[h]
\psfig{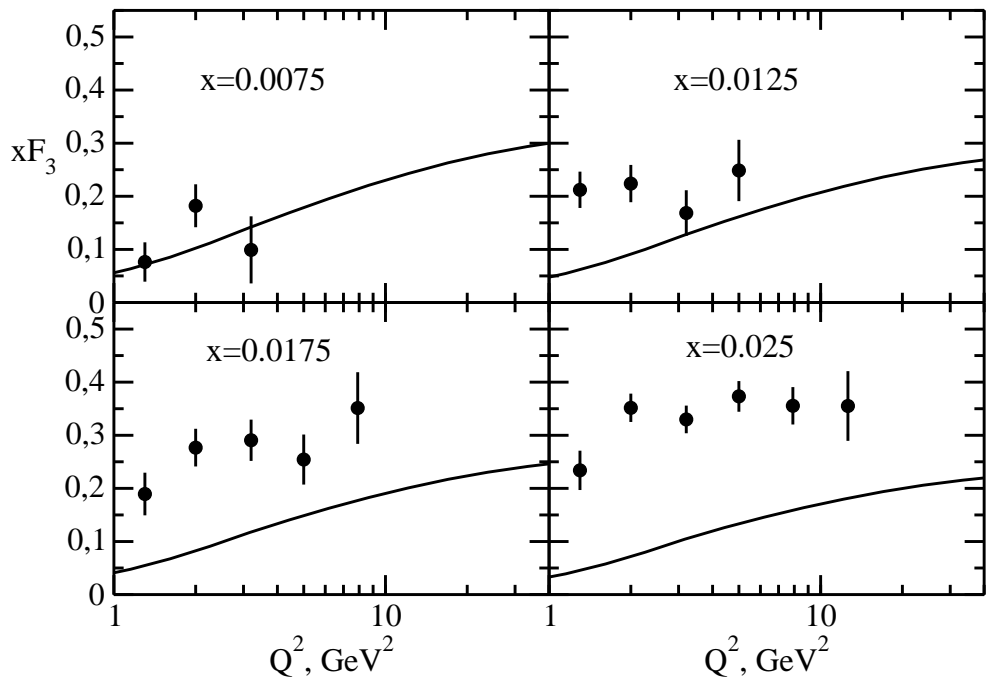}
\vspace{-0.5cm}
\caption{Data points are CCFR measurements of $xF_3(x,Q^2)$ \cite{CCFR}. 
Curves show the vacuum exchange contribution to  $xF_3(x,Q^2)$.} 
\label{fig:fig1}
\end{figure}  

We evaluate $xF_3(x,Q^2)$  making use of Eqs.~(\ref{eq:FACTOR}) and 
(\ref{eq:SIGMA}) with the  differential
gluon density function
${{\cal F}(x_g,\kappa^2)}$ determined in \cite{NIKIV}. As reported in
\cite{NIKIV}, the approach
developed  
works very well in the perturbative region
of high $Q^2$ and small $x~ (x\lsim 0.01)$. Besides, a realistic extrapolation of 
${{\cal F}(x_g,\kappa^2)}$ into
the soft region allows calculations
 at lowest $Q^2$ also \cite{NIKIV}. 
In our calculations for $Q^2\lsim M^2=2(m^2+\mu^2)$
the gluon density 
${{\cal F}(x_g,\kappa^2)}$ 
 enters Eq.~(\ref{eq:SIGMA}) at the gluon momentum fraction
$x_g=x(1+M^2/Q^2)$. For large virtualities,  $Q^2\gsim  M^2$, 
we put  $x_g= 2x$.
Direct evaluation of the proton DIS structure function $F_{2p}(x,Q^2)$
shows that this prescription  corresponding to the
collinear DLLA ensures a good description of
experimental data on the  light and heavy flavor electro-production
in a wide  range of the photon virtualities down to $Q^2\sim 1$ GeV$^2$.
   The constituent
quark masses are as follows  $m_u=m_d=0.2$ GeV, $m_s=0.35$ GeV  and 
$m_c=1.3$ GeV.

 The  $xF_3$ data reported by the CCFR Collaboration are presented 
 in Figure \ref{fig:fig1}.  Shown is  the $Q^2$-dependence 
   of    $xF_3$  for several  smallest values of $x$ \cite{CCFR}. 
It should be
emphasized that we focus on  the vacuum exchange contribution 
to $xF_3$ corresponding
to the excitation of the $c\bar s$ state in the process (\ref{eq:WG}).
Therefore, the structure function $xF_3$ 
differs from zero only  due to the strong left-right asymmetry of the 
light-cone $|c\bar s\rangle$ Fock state. 
Shown by the solid line in Fig. \ref{fig:fig1} is the Pomeron exchange
 contribution to $xF_3$. The latter can be interpreted in terms 
of parton densities  as the sea-quark component 
of $xF_3$.  

\begin{figure}[h]
\psfig{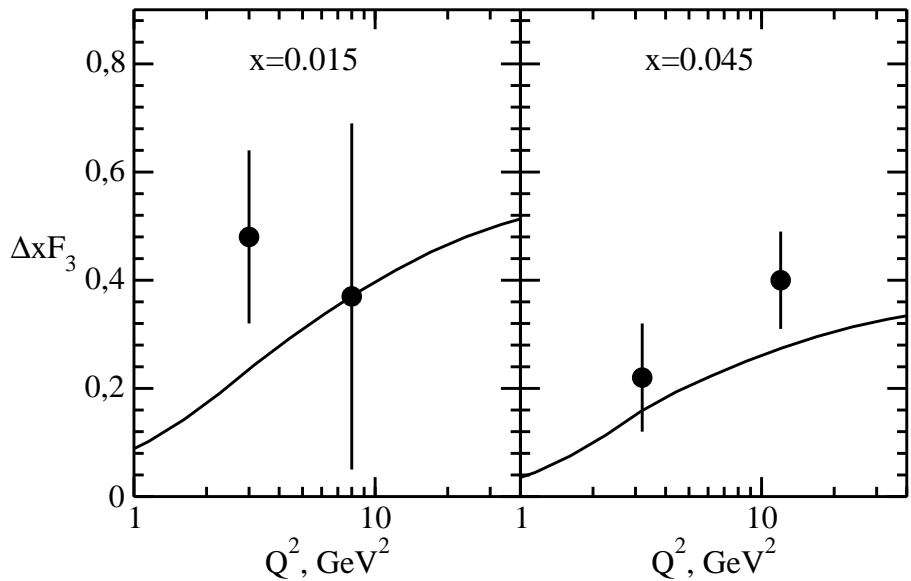}
\caption{$\Delta x F_3$ data  as a function of $Q^2$ \cite{CCFR3}.
 Shown by  solid lines are the results of  color dipole description.} 
\label{fig:fig2}
\end{figure}
Looking at Figure \ref{fig:fig1}
one should bear in mind that the smallest available values of $x$ 
are in fact only moderately small and there is also 
quite significant  valence contribution to $xF_3$.
 The valence term, $xV$ is the same
for both  $\nu N$ and 
$\bar\nu N$ structure functions of an  iso-scalar nucleon. 
The sea-quark term in the $xF^{\nu N}_3$ denoted by $xS(x,Q^2)$ has opposite
sign for $xF^{\bar\nu N}_3$, the substitution 
$m\leftrightarrow\mu$ in Eqs.~(\ref{eq:RHOR}) and (\ref{eq:RHOL}) entails
$\sigma_L\leftrightarrow\sigma_R$.
 Therefore,  
\beq
xF^{\nu N}_3=xV+xS,
\label{eq:SVNU}
\eeq
and
\beq
xF^{\bar\nu N}_3=xV-xS.
\label{eq:SVANU}
\eeq
One can combine the $\nu N$ and 
$\bar\nu N$ structure functions to isolate the Pomeron exchange term,
\beq
\Delta x F_3=xF_3^{\nu N}-xF_3^{\bar \nu N}=2xS.
\label{eq:fig2}
\eeq
The extraction of $\Delta x F_3$
from CCFR $\nu_{\mu}Fe$ and $\bar\nu_{\mu}Fe$ differential cross section
in a model-independent way has been reported in \cite{CCFR3}. Figure 
\ref{fig:fig2} shows  the extracted values of $\Delta x F_3$
as a function of  $Q^2$  for two 
smallest values of $x$. Also shown are the results of our calculations.

After evaluating the difference of left and right cross sections
let us turn to  their sum and, as a consistency check, 
evaluate  the structure function
\beq
2xF_1(x,Q^2)={Q^2\over 4\pi^2\alpha_{W}}\sigma_{T}(x,Q^{2}),
\label{eq:F1}
\eeq
where
\beq
\sigma_{T}=
{1\over 2}
\left[\sigma_{L}(x,Q^{2})+\sigma_{R}(x,Q^{2}) \right].
\label{eq:SIGT}
\eeq
\begin{figure}[h]
\psfig{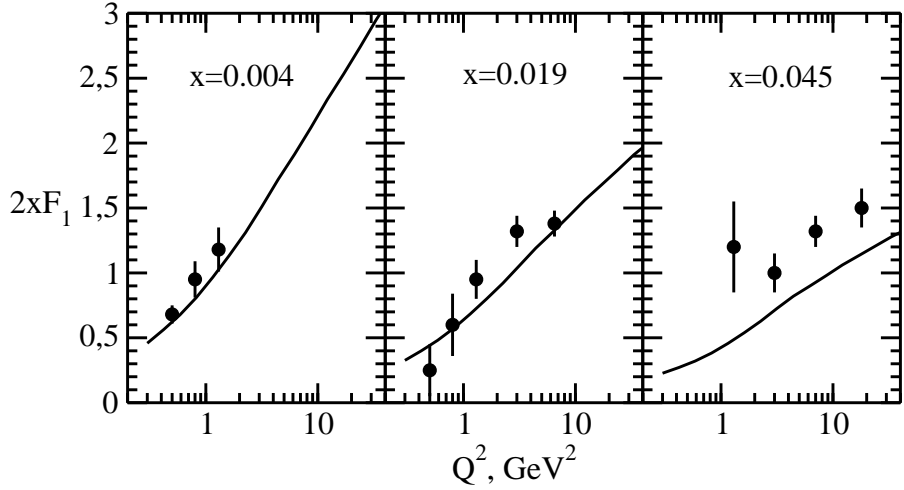}
\vspace{-0.5cm}
\caption{CCFR measurements of $2xF_1(x,Q^2)$  \cite{CCFR2} 
compared with our estimates. Curves show the vacuum  exchange contribution to
 $2xF_1(x,Q^2)$.} 
\label{fig:fig3}
\end{figure}
The CCFR Collaboration  measurements \cite{CCFR2} of 
the structure function  $2xF_1$ as a function of $Q^2$ 
for three values of $x$ 
are shown in Fig. \ref{fig:fig3}. Theory and experiment here  are in  
qualitatively the same relations as in Fig. \ref{fig:fig1}. 
In small-$x$ region, $x<0.01$,
dominated by the Pomeron exchange  our estimates are  in agreement
with data. For larger $x$ the non-vacuum  contributions enter the game 
 and a certain divergence shows up. 
 This divergence will
increase if we take into account the nuclear effects.
Indeed, the CCFR/NuTeV structure functions 
$xF^{\nu N}_3$ and $xF^{\bar\nu N}_3$ are  extracted from the 
$\nu Fe$  and $\bar\nu Fe$ data.
The nuclear thickness factor, $T(b)=\int dz n(\sqrt{z^2+b^2})$, 
where $b$ is the impact parameter and $n(r)$ is the nuclear matter density,
$\int d^3r n(r)=A$,
makes the nuclear cross section  
\beq
\sigma^A_{\lambda}=
A\langle\sigma_{\lambda}\rangle
-\delta\sigma^A_{\lambda},
\label{eq:NUCLEAR}
\eeq
with the nuclear shadowing term 
\beq
\delta\sigma^A_{\lambda}\simeq
{\pi\over 4} \langle \sigma_{\lambda}^2\rangle
\int db^2 T(b)^2~,
\label{eq:SHADOW}
\eeq
very sensitive to the left-right asymmetry of the $\nu$-nucleon 
cross sections.
In Eqs.~(\ref{eq:NUCLEAR}) and (\ref{eq:SHADOW})
$
\langle\sigma_{\lambda}\rangle=
\langle\Psi_{\lambda}|\sigma(x,r)  |\Psi_{\lambda} \rangle
$
and
$
\langle \sigma_{\lambda}^2\rangle=
\langle\Psi_{\lambda}|\sigma(x,r)^2|\Psi_{\lambda} \rangle.
$
 Hence, the nuclear shadowing correction  
\bea
\delta xF_3\simeq {Q^2\over 4\pi^2\alpha_W}
{\pi\langle \sigma_L^2-\sigma_R^2\rangle\over 8A}
\int db^2 T(b)^2,
\label{eq:DELF3}
\eea
which should be added to  $xF_3$ extracted from
the $\nu Fe$ data to get the  ``genuine'' $xF_3$. 
 Since $\langle \sigma_L^2\rangle\propto 1/\mu^2$ and
$\langle\sigma_R^2\rangle\propto 1/m^2$,  this correction is large,
 positive-valued  and does
increase $xF_3$ of the impulse approximation.

Summarizing, we 
developed the light-cone color dipole
description  of the left-right asymmetry effect 
in charged current DIS at small Bjorken $x$. 
We compared our results with experimental data  and found  
a considerable vacuum exchange contribution to the structure functions 
$xF^{\nu N}_3$. This contribution is found to dominate
 the structure function 
$\Delta x F_3=xF_3^{\nu}-xF_3^{\bar \nu}$ of  an  iso-scalar nucleon
extracted from  nuclear data. Theory is in reasonable agreement with data
but the nuclear effects are shown can make this comparison
a somewhat more complicated procedure.
The color dipole analysis of nuclear effects in the CC DIS 
will be published elsewhere.
 
\vspace{0.2cm} \noindent \underline{\bf Acknowledgments:}
V.R.~Z. thanks  the Dipartimento di Fisica dell'Universit\`a
della Calabria and the Istituto Nazionale di Fisica
Nucleare - gruppo collegato di Cosenza for their warm
hospitality while a part of this work was done.
The work was supported in part by the Ministero Italiano
dell'Istruzione, dell'Universit\`a e della Ricerca.

\end{document}